\documentclass[twocolumn,amsmath,amssymb,10pt,superscriptaddress,a4paper,letterpaper,fleqn]{revtex4-1}
\usepackage{amssymb}
\usepackage{epsfig}
\usepackage{graphicx}
\usepackage{dcolumn}
\usepackage{array}
\usepackage{bm}
\usepackage{fancyhdr}
\usepackage{longtable}
\usepackage{multirow}
\usepackage{float}
\usepackage{afterpage}
\usepackage{color}

\setlongtables
\usepackage[breaklinks=true,linkbordercolor={1 1 1}]{hyperref}

\begin{document}
\setcounter{page}{1}

%%% **********************************************************************

\title{
%% Please do not remove the line below
\qquad \\ \qquad \\ \qquad \\  \qquad \\  \qquad \\ \qquad \\
%% Change title, authors, afiliation and type  your abstract
Sensitivity of measured fission yields on prompt-neutron corrections}

\author{A. Al-Adili}
\email[]{Ali.Al-Adili@physics.uu.se}
\affiliation{Division of Applied Nuclear Physics, Uppsala University, S-751 20 Uppsala, Sweden}
\affiliation{European Commission - DG Joint Research Centre (IRMM), B-2440 Geel, Belgium}

\author{F.-J. Hambsch}
\email[]{Franz-Josef.HAMBSCH@ec.europa.eu}
\affiliation{European Commission - DG Joint Research Centre (IRMM), B-2440 Geel, Belgium}

\author{S. Pomp}
\affiliation{Division of Applied Nuclear Physics, Uppsala University, S-751 20 Uppsala, Sweden}

\author{S. Oberstedt}
\affiliation{European Commission - DG Joint Research Centre (IRMM), B-2440 Geel, Belgium}

\date{\today} 
%\received{8 March 2013; revised received XX June 2013; accepted XX September 2013}

\begin{abstract}
{
The amount of emitted prompt neutrons from the fission fragments increases as a function of excitation energy. 
Yet it is not fully understood whether the increase in $\bar{\nu} \left( A \right)$ as a function of $E_\mathrm{n}$ is mass dependent. The share of excitation energies among the fragments is still under debate, but there are reasons to believe that the excess in neutron emission originates only from the heavy fragments, leaving $\bar{\nu}_\mathrm{light} \left( A \right)$ almost unchanged. In this work we investigated the consequences of a mass-dependent increase in $\bar{\nu} \left( A \right)$ on the final mass and energy distributions. The assumptions on $\bar{\nu} \left( A \right)$ are essential when analysing measurements based on the 2E-technique. This choice showed to be significant on the measured observables. For example, the post-neutron emission mass yield distribution revealed changes up to 10-30 \%. The outcome of this work pinpoint the urgent need to determine $\bar{\nu} \left( A \right)$ experimentally, and in particular, how $\bar{\nu} \left( A \right)$ changes as a function of incident-neutron energy. Until then, many fission yields in the data libraries could be largely affected, since they were analysed based on another assumption on the neutron emission. 
}
\end{abstract}
\maketitle

%%% DO NOT EDIT the following section enclosed by *****
%%% ***************************************************
%\lhead{ND 2013 Article $\dots$}
%\chead{NUCLEAR DATA SHEETS}
\rhead{A. Al-Adili, F.-J. Hambsch, S. Pomp and S. Oberstedt }
\lfoot{}
\rfoot{}
\renewcommand{\footrulewidth}{0.4pt}
%%% ***************************************************

%%% EDIT: the body of your text starts here, you may use as many \section, \subsection, \subsubsection
%%% \begin{figure}, \begin{tabular} and \begin{equations} as needed. Please note that each \begin{}
%%% must be closed with the corresponding \end{} and that section titles should be in capital
%%% letters. Current text should be eventually deleted.

\section{Introduction}
Accurate fission-fragment (FF) yield measurements are essential both for reactor- and model calculations. They are needed in various stages of the nuclear fuel cycle,  e.g. reactor criticality- and safety calculations, burnup determination, decay-heat calculations and processing of spent nuclear fuel \cite{IAEA1168}. Fission yields are also important for the development of fission models. Countless yield measurements have been performed utilizing different techniques. This study focuses on the widely used double energy technique (2E) in which the mass yields are determined by measuring the FF kinetic energies. This technique suffers from the vital need to assume the neutron emission on an event-by-event basis. In this work we show that this assumption has a rather strong implication on the measured yields. 

The accelerated FF remove the major part of their excitation energies by emitting prompt neutrons. The prompt-neutron multiplicity $\bar{\nu}$ depends on the fragment mass, kinetic energy and excitation energy in the fissioning system. The mass dependence is governed by the deformation of the fragments and by shell effects. Due to the doubly magic nucleus at \textit{A} = 132, the neutron emission is highly suppressed. The resulting $\bar{\nu} \left( A \right)$ shape has a "saw-tooth" like structure and is shown for instance in Figs. \ref{fig1}(a,b) \cite{naqvi,muller}. Measurements of the total average neutron multiplicity ($\bar{\nu}_{\mathrm{tot}}$), e.g. Ref. \cite{mather} for $^{234} \mathrm{U} \left (n,f \right)$, showed an increasing trend as a function of $E_\mathrm{n}$. At $E_\mathrm{n} \approx$ 0 MeV, $\bar{\nu} \left( A \right)$ is roughly 2.4 and it becomes roughly 3.0 at $E_\mathrm{n}$ = 5 MeV. In various models and experiments, these extra neutrons are assumed to be independent on the fragment mass \cite{vives,siegler,Aliphd,streade,lestone,yong,vogt}. Therefore, $\bar{\nu} \left( A \right)$ was increased almost equally for the light and heavy fragments. However some measurements found that the extra neutrons are predominantly emitted from the heavy fragments, e.g. in $^{237} \mathrm{Np} \left (n,f \right)$ \cite{naqvi} and $^{235} \mathrm{U} \left (n,f \right)$ \cite{muller} (see Figs. \ref{fig1}(a,b)). These experiments used the (2E-2v) technique which is not confined by the neutron emission assumption since the FF velocities are also measured. Both pre- and post neutron-emission distributions were determined, which allowed for an estimation of $\bar{\nu} \left( A \right)$. Recently, theoretical models explained this increase in $\bar{\nu}_{\mathrm{heavy}}\left( A \right)$ based on the energy-sorting mechanism \cite{schmidt}. In the pre-scission state, the fragments are still connected through the neck and can exchange excitation energy. The temperature is defined as the mean excitation energy per degree-of-freedom, and depends on the fragment mass as $T\propto A^{-2/3}$ \cite{Egidy}. The light fragments have higher temperatures than the heavy ones. Because of the thermal contact, the energy will flow to the heavy fragments. Due to pairing correlations the nucleons rearrange so that the temperature stays constant during the entire process, despite the added excitation energy to the heavy fragment. The energy sorting is believed to behave similarly for neighbouring nuclei. The GEF code uses the energy sorting and can be used to estimate the expected behaviour \cite{GEF}. A GEF calculation was performed for $^{234} \mathrm{U} \left (n,f \right)$ and can be seen in Fig. \ref{fig1}(c). 

\section{The 2E-technique}
The present work aims to study the sensitivity of the measured yield distributions on the assumption of $\bar{\nu} \left( A \right)$. Experimental data on $^{234} \mathrm{U} \left( n,f \right)$ were used at $E_\mathrm{n}$ = 4 and 5 MeV \cite{Aliphd}. It is still argued whether the change of $\bar{\nu} \left( A \right)$ at different $E_\mathrm{n}$ is mass-dependent. However if the neutron emission proves to increase only from the heavy fragments, the measured yields based on the 2E-technique, might be affected if they have been calculated using the average increase model. In the 2E-technique both FF kinetic energies are measured in combination usually with their emission angles relative to the incoming beam. Conservation of momentum and mass are used to iteratively calculate the pre-neutron emission masses (for details on the kinematics see Ref. \cite{Aliphd}). The detected post-neutron energies ($E_{\mathrm{post}}^{\mathrm{LAB}}$) are converted to pre-neutron energies ($E_{\mathrm{pre}}^{\mathrm{LAB}}$) via:
%---------------------- Eq
\begin{equation}
\label{Eqkinem7}
E_{\mathrm{pre}}^{\mathrm{LAB}} \approx  E_{\mathrm{post}}^{\mathrm{LAB}} A_{\mathrm{pre}} \times \left( A_{\mathrm{pre}}-\nu \left( A,\mathrm{TKE},E_\mathrm{n} \right) \right)^{-1} \quad.
\end{equation}
%---------------------- 
%-//FIG------------------------
\begin{figure}[p]
\begin{center}
\includegraphics[width=8.4cm]{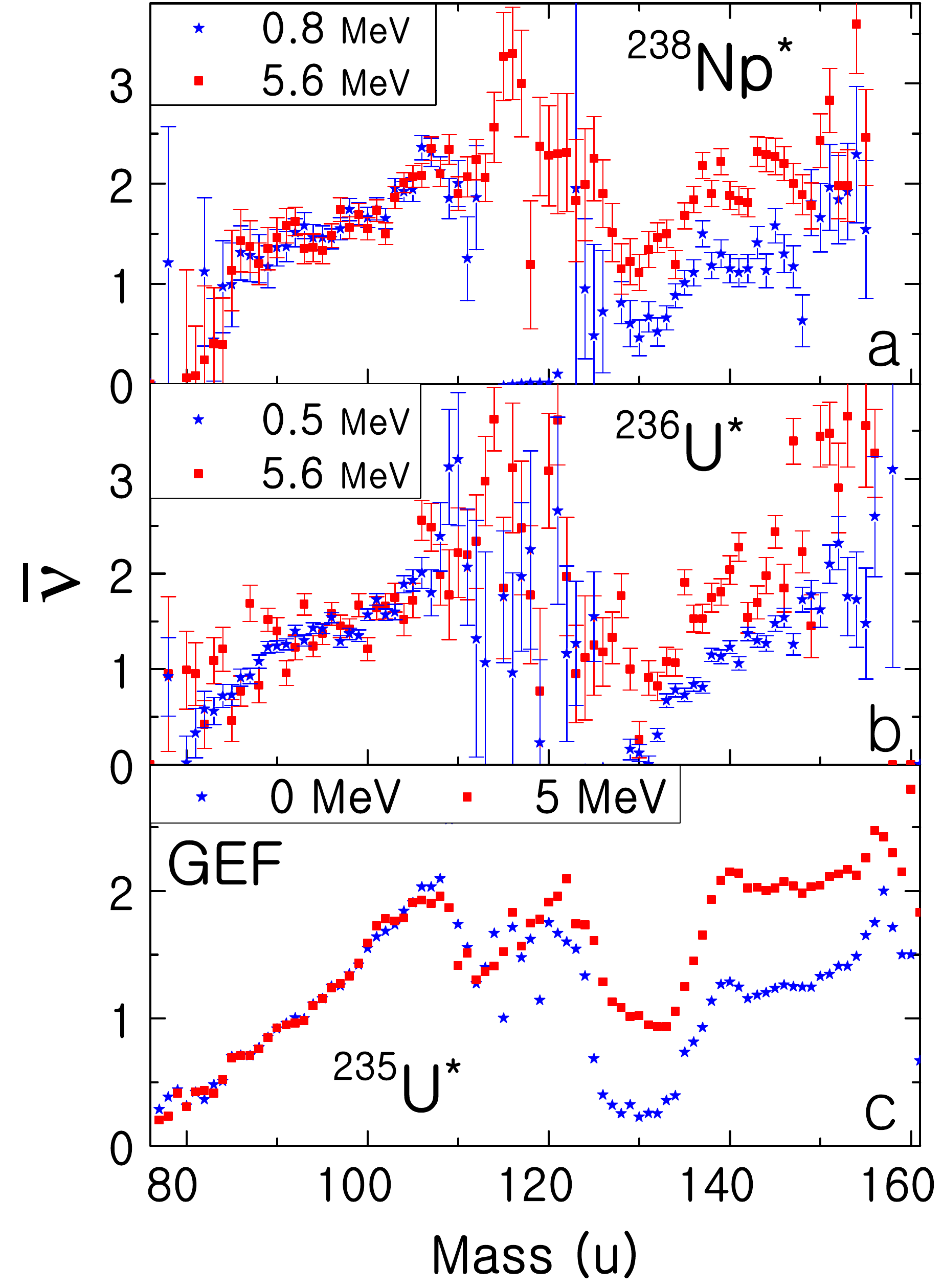}
\caption{\label{fig1} (a) $\bar{\nu} \left( A \right)$ for $^{237}\mathrm{Np}\left( n,f \right)$ at different $E_\mathrm{n}$ \cite{naqvi}. (b) $\bar{\nu} \left( A \right)$ for $^{235}\mathrm{U}\left( n,f \right)$ at different $E_\mathrm{n}$ \cite{muller}. (c) GEF calculation performed for $^{234} \mathrm{U} \left (n,f \right)$ at $E_\mathrm{n}$ = 0 and 5 MeV.}

\includegraphics[width=8.4cm]{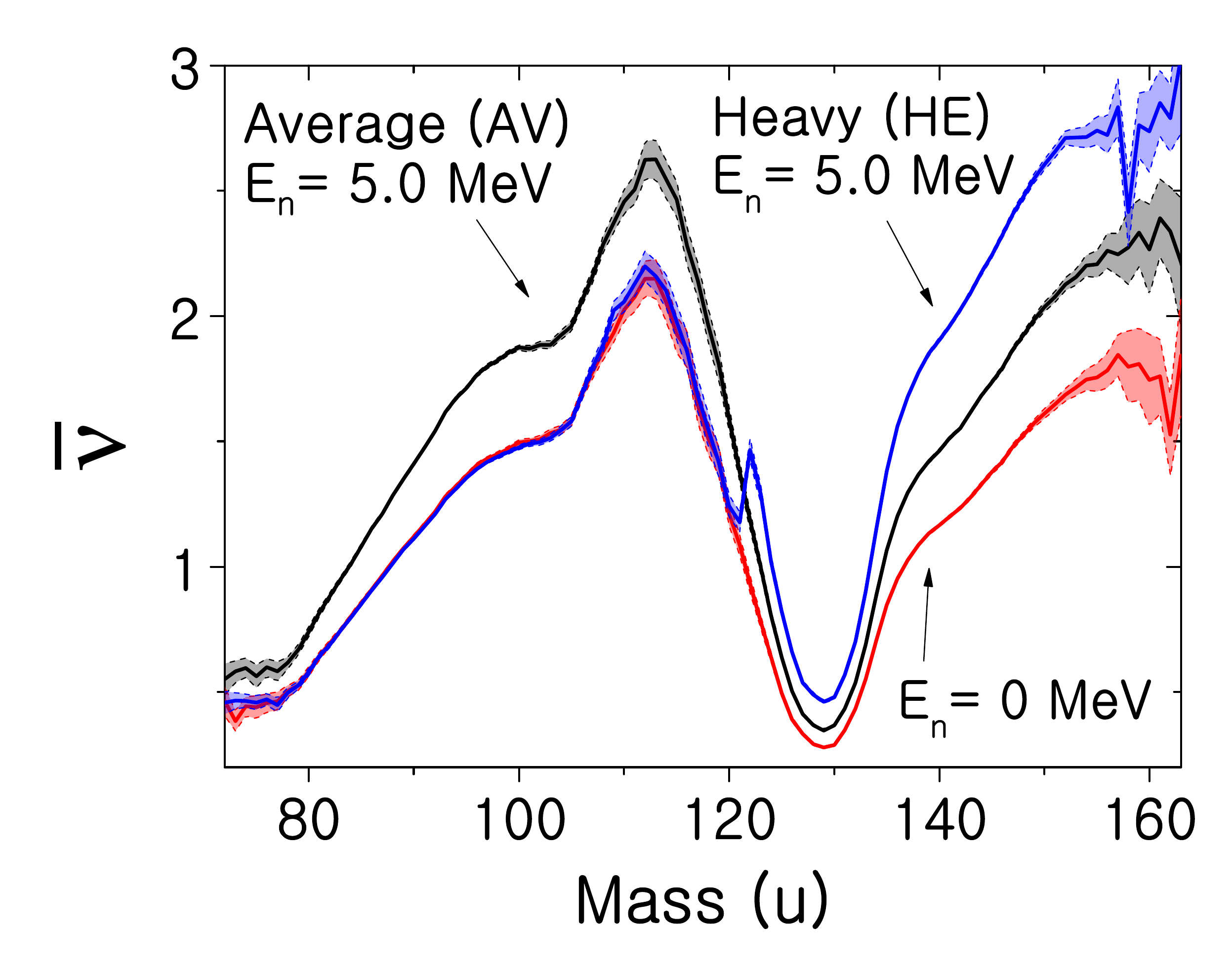}
\caption{\label{fig2} $\bar{\nu}_{234} \left( A \right)$ for $E_\mathrm{n}$ = 0 and 5 MeV deduced from the average of $^{233,235}$U \cite{wahl}. The original $\bar{\nu}_{234} \left( A \right)$ shape was multiplied by a factor ($\alpha$) and weighted with the mass distribution to give $\bar{\nu}_{\mathrm{tot}}$ at $E_\mathrm{n}$ = 5 MeV. Two different assumptions were compared for the increase in $\bar{\nu}$ as a function of $E_\mathrm{n}$. The first using $\alpha_1$ acting over all masses, the second using $\alpha_2$ acting on the heavy fragments only ($A>120$). In order to yield the same $\bar{\nu}_{\mathrm{tot}}$ at $E_\mathrm{n}$ = 5 MeV, $\alpha_2>\alpha_1$.}

\end{center}
\end{figure}
%-//FIG------------------------
In this approximation the recoil from the neutron emission to the accelerated fragments is neglected. In most cases, the neutron emission is not measured so an approximation is needed. On average, the neutrons are isotropically emitted in the CM system.  Moreover the added recoil is very small due to the mass difference of the neutron and the fragment. Unless measuring the neutron emission on an event-by-event basis, these assumptions are unavoidable even when using the 2E-2v method. The $E_{\mathrm{post}}^{\mathrm{LAB}}$ are converted into the centre-of-mass system and used to calculate the pre-neutron emission masses: 
 %---------------------- Eq
\begin{equation}
\label{Eqkinem10}
A_{1,2} = A_{\mathrm{CN}} E^{\mathrm{CM}}_{2,1} \times \left( E^{\mathrm{CM}}_1+E^{\mathrm{CM}}_2 \right)^{-1} \quad,
\end{equation}
%----------------------
were (1,2) denote the two FF. The neutron multiplicity is updated in each iteration based on the newly determined masses and energies. The neutron emission has to be parametrized as a function of mass and TKE. Moreover, the total neutron emission has to correspond to $\bar{\nu}_{\mathrm{tot}}$ from Ref. \cite{mather}. Two approaches (see Fig. \ref{fig2}) of the correction were tested individually and compared side-by-side:
%-//FIG------------------------
\begin{figure*}[]
\begin{center}
\includegraphics[width=\textwidth]{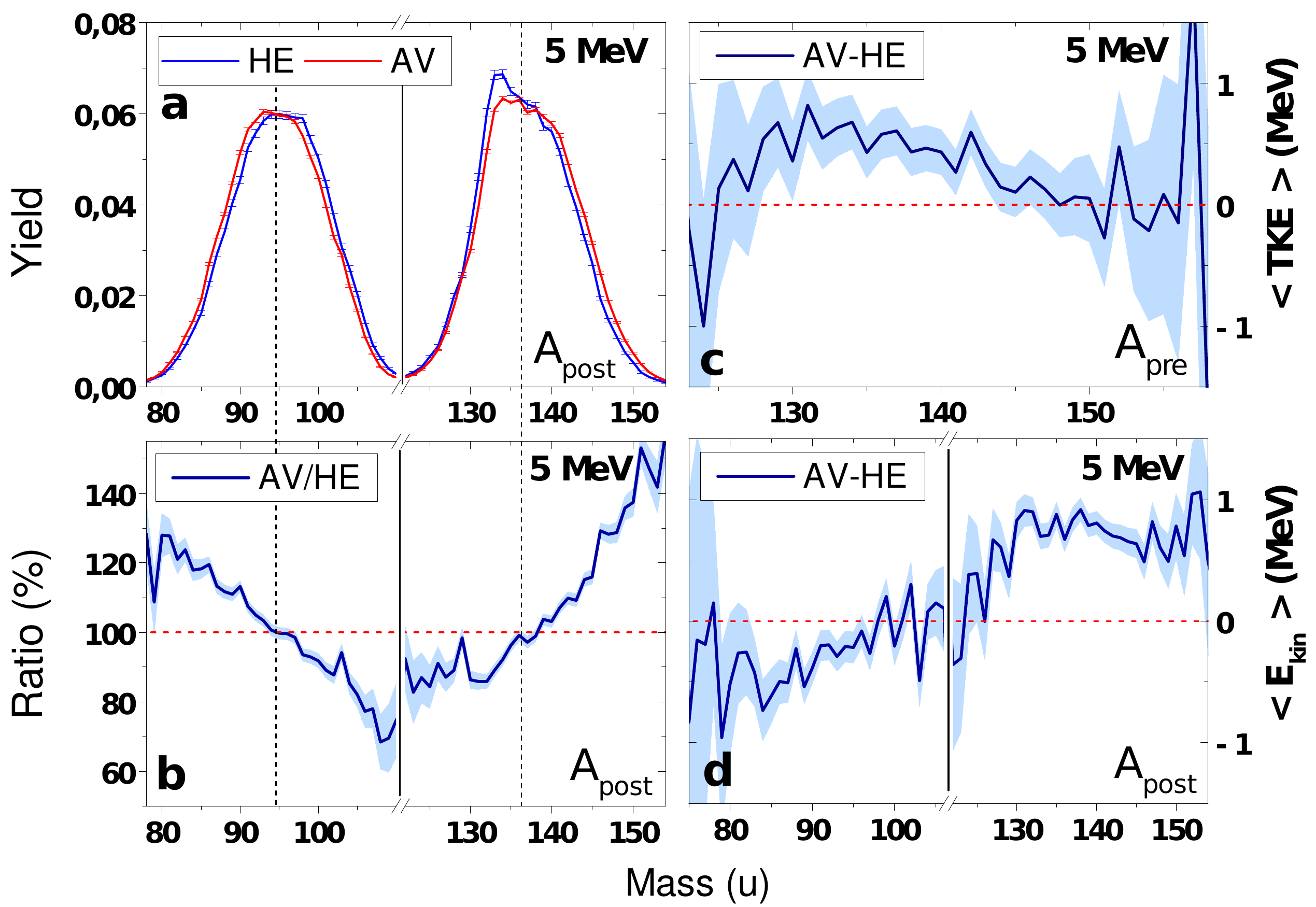}
\caption{\label{fig3}(a) The post-neutron mass distribution for $^{234} \mathrm{U} \left( n,f \right)$ at $E_ \mathrm{n}$ = 5 MeV, for both applied corrections (AV and HE). (b) The ratio between the distributions. (c) The difference in $\langle \mathrm{TKE} \rangle$ as a function of pre-neutron emission mass. (d) The difference in $\langle E_\mathrm{post} \rangle$ as a function of post-neutron emission mass. Only statistical errors are shown.}
\end{center}
\end{figure*}
%-//FIG------------------------
\begin{enumerate}
\item[$\diamondsuit$]Average method (AV): $\bar{\nu}\left( A \right)$ was assumed to increase for all masses. The original $\bar{\nu}\left( A \right)$ was multiplied by a factor ($\alpha_1$) and adjusted (weighted with the mass yields) to correspond to the expected $\bar{\nu}_{\mathrm{tot}}$. 

 \item[$\diamondsuit$]Heavy method (HE):  $\bar{\nu}\left( A \right)$ increases for $A >$ 120. The original sawtooth curve for the heavy masses was multiplied with a factor ($\alpha_2$) to give the total expected neutron emission. For $A \leq$ 120, the distribution was kept unchanged at $\bar{\nu}_{\mathrm{th}} \left( A \right)$.
\end{enumerate}

When applying the AV method e.g. on the 5 MeV case, both the light and heavy fragments get an increase of roughly 0.3 neutrons. Based on the HE method instead, the heavy fragments emit all extra 0.6 neutrons. In both methods, $\alpha$ is fine-tuned to give $\bar{\nu}_{\mathrm{tot}} \approx$ 3 neutrons.

\section{Impact of the different assumptions}
The difference in neutron emission per fragment had a clearly visible impact on the distributions. The two mass peaks moved closer to each other by using the HE-method instead of the AV-method. This was observed both in the pre-neutron emission and the post-neutron emission distributions. However the effects were at least twice as pronounced in the latter case. The impact is indirectly affecting the pre-neutron masses during the iteration calculations. But it directly affects the post-neutron masses since those are determined by a direct subtraction $A_{\mathrm{post}}=A_{\mathrm{pre}}-\nu$. Fig. \ref{fig3} shows some examples of the effects observed. In (a) the post-neutron mass distributions for $E_\mathrm{n}$ = 5 MeV are plotted. The distributions show an overall shift and a small change in the heavy peak width. Specific masses experience large differences, especially around $A \approx$ 132. The ratio between the two distributions is plotted in Fig. \ref{fig3}(b). As seen, the differences are about 10-15 \% for masses with 0.04 absolute yield. For masses with 0.03 absolute yield, the differences reach between 20-30 \%. These large discrepancies will have a considerable influence on reactor- and model calculations. The accuracy often requested is far below these levels. Furthermore, it was found that all effects increase with the incident-neutron energy therefore larger corrections are needed when using yields at higher excitation energies. The energy distribution is also affected as seen in Fig. \ref{fig3}(c). The change in TKE distribution reach up to $ \Delta \mathrm{TKE} =$ 0.5 MeV for masses around $A$ = 132. For higher asymmetries, the differences are compensated between the AV and HE distributions. The single fragment kinetic energies are plotted in Fig. \ref{fig3}(d) as a function of $A_{\mathrm{post}}$. The heavy fragments overall have larger kinetic energy when using the AV-method. The light fragments change however differently: The HE-method gives higher kinetic energies but most notably at higher asymmetry. The observed changes are obviously mass-dependent which increases the complexity in the needed corrections. For the full details on the observed differences, see Ref. \cite{AliPRC}.

The average changes in TKE and mass distribution were studied. The TKE obtained from the HE-method is smaller than the AV-method and can be estimated by: 
%- EQ4 ------------------------------------------------------------------
\begin{equation}
\mathrm{ \overline{TKE} _{HE}} \approx \mathrm{ \overline{TKE} _{AV}}-0.038\times E_\mathrm{n}\; \;(\mathrm{MeV}) 
\label{tkecorr}
\end{equation}
%-------------------------------------------------------------------------------
The pre-neutron mass distribution changes roughly by:
%- EQ5 ---------------------------------------------------------------------
\begin{equation}
\langle A^{\mathrm{pre}}_\mathrm{H}  \rangle _{\mathrm{HE}} \approx \langle A^{\mathrm{pre}}_\mathrm{H}  \rangle _{\mathrm{AV}}-0.065\times E_\mathrm{n}\; \;(\mathrm{u})
\label{mprecorr}
\end{equation}
%-------------------------------------------------------------------------------
The post-neutron mass distributions showed double the effect compared to the pre-neutron case. At $E_\mathrm{n}$ = 5 MeV, the change in average mass is about 0.7 amu:
%- EQ6 --------------------------------------------------------------------
\begin{equation}
\langle A^{\mathrm{post}}_\mathrm{H} \rangle _{\mathrm{HE}} \approx \langle A^{\mathrm{post}}_\mathrm{H}  \rangle _{\mathrm{AV}}-0.135\times E_\mathrm{n}\; \;(\mathrm{u})
\label{mpostcorr}
\end{equation}
%--------------------------------------------------------------
Eqs. (\ref{tkecorr}, \ref{mprecorr}, \ref{mpostcorr}) may be used as a first correction to the expected shift. However, average shifts are probably not sufficient to account for the total changes, since the differences seen in Fig. \ref{fig3}(a-d) are mass-dependent. 

\section{Conclusions}
In this work, we investigated the impact of different $\bar{\nu}\left( A \right)$ corrections on the FF distributions from the $^{234} \mathrm{U} \left( n,f \right)$ reaction at $E_\mathrm{n}$ = 4 and 5 MeV. The increase of $\bar{\nu}\left( A \right)$ as a function of $E_\mathrm{n}$ was under focus and it was questioned how $\bar{\nu}\left( A \right)$ changes. Two assumptions were used, the first increasing $\bar{\nu}\left( A \right)$ for all fragments, the second increasing $\bar{\nu}\left( A \right)$ for the heavy fragments only. The impact found was large on both the yield- and energy distributions. The differences are mass-dependent and they grew as a function of $E_\mathrm{n}$. The post-neutron emission mass yields were affected by up to 20-30 \% for certain masses. Many fission yields in the data libraries were measured via the 2E-technique assuming an average increase of $\bar{\nu}\left( A \right)$. If the HE-method proves valid, all these yields need to be corrected. Hence, this study emphasizes the need of determining $\bar{\nu} \left( A \right)$ as a function of mass and $E_\mathrm{n}$. 

\section{Outlook}
The important remaining question is how the extra neutron emission depends on the fragment mass. The Uppsala group and the JRC-IRMM are involved in three different projects which may contribute to solving this issue:

\begin{enumerate}
\item[$\diamondsuit$]Fission yields at IGISOL: Currently, the Uppsala group is designing a mono-energetic neutron source for the IGISOL-JYFLTRAP facility. The independent fission products will be identified via an ion-guide technique and a Penning trap, providing a high-precision A/q determination. The aim is to study the yield distribution as a function of $E_\mathrm{n}$ and to estimate the change of $\bar{\nu} \left( A \right)$ \cite{Solders}.\\

\item[$\diamondsuit$]Coincidence measurements: The JRC-IRMM is developing a neutron-detector array to measure prompt-fission neutrons in coincidence with FF via an ionisation chamber \cite{Hambsch}. It is straight forward then to  measure at different $E_\mathrm{n}$ and study how $\bar{\nu} \left( A \right)$ changes for the light and heavy fragments. 

\item[$\diamondsuit$]The VERDI spectrometer:  At the JRC-IRMM, a double (v, E)
fission-fragment time-of-flight spectrometer (VERDI) is under development \cite{Oberstedt}. By measuring both the FF velocities and energies, one can estimate $\bar{\nu} \left( A \right)$ on an event-by-event basis. 
 
 \end{enumerate}

%%% IMPORTANT: When preparing bibliography observe strictly the layout below (initials in front of the names,
%%% no names in capitals, not more than four authors, no titles, volume number in bold, year at the end within parenthesis,
%%% dot (.) at the end.

\section*{Acknowledgements}
A. A. would like to thank the European Commission, Joint Research Centre, for granting him a Ph.D. fellowship at the JRC-IRMM.

\end{document}